\documentclass[]{aastex631}

\shorttitle{Marked power spectrum of the galaxy field}
\shortauthors{Massara et al.}

\begin{document}

\title{Cosmological Information in the Marked Power Spectrum of the Galaxy Field}

\author[0000-0002-0637-8042]{Elena Massara}
\correspondingauthor{Elena Massara}
\email{elena.massara.cosmo@gmail.com}
\affiliation{Waterloo Centre for Astrophysics, University of Waterloo, 200 University Ave W, Waterloo, ON N2L 3G1, Canada }
\affiliation{Department of Physics and Astronomy, University of Waterloo, Waterloo, ON N2L 3G1, Canada}

\author{Francisco Villaescusa-Navarro}
\affiliation{Center for Computational Astrophysics, Flatiron Institute, 162 5th Avenue, New York, NY 10010, USA}
\affiliation{Department of Astrophysical Science, Princeton University, Peyton Hall, Princeton NJ 08544, USA}

\author{ChangHoon Hahn}
\affiliation{Department of Astrophysical Science, Princeton University, Peyton Hall, Princeton NJ 08544, USA}

\author{Muntazir M. Abidi}
\affiliation{D\'epartement de Physique Th\'eorique, Universit\'e de Gen\`eve, 24 quai Ernest Ansermet, 1211 Gen\`eve 4, Switzerland}

\author{Michael Eickenberg}
\affiliation{Flatiron Institute Center for Computational Mathematics, 162 5th Ave, 3rd floor, New York, NY 10010, USA}

\author{Shirley Ho}
\affiliation{Center for Computational Astrophysics, Flatiron Institute, 162 5th Avenue, New York, NY 10010, USA}
\affiliation{Department of Astrophysical Science, Princeton University, Peyton Hall, Princeton NJ 08544, USA}
\affiliation{Center for Cosmology and Particle Physics, Department of Physics, New York University, New York, NY 10003, USA}
\affiliation{Department of Physics, Carnegie Mellon University, Pittsburgh, PA 15217, USA}

\author{Pablo Lemos}
\affiliation{Department of Physics and Astronomy, University of Sussex, Sussex House, Falmer, Brighton, BN1 9RH, UK}
\affiliation{Department of Physics and Astronomy, University College London, Gower Street, London, WC1E 6BT, UK}

\author{Azadeh Moradinezhad Dizgah}
\affiliation{D\'epartement de Physique Th\'eorique, Universit\'e de Gen\`eve, 24 quai Ernest Ansermet, 1211 Gen\`eve 4, Switzerland}

\author{Bruno R\'egaldo-Saint Blancard}
\affiliation{Flatiron Institute Center for Computational Mathematics, 162 5th Ave, 3rd floor, New York, NY 10010, USA}



\begin{abstract}
Marked power spectra are two-point statistics of a marked field obtained by weighting each location with a function that depends on the local density around that point. We consider marked power spectra of the galaxy field in redshift space that up-weight low density regions, and perform a Fisher matrix analysis to assess the information content of this type of statistics using the Molino mock catalogs built upon the Quijote simulations. We identify four different ways to up-weight the galaxy field, and compare the Fisher information contained in their marked power spectra to the one of the standard galaxy power spectrum, when considering monopole and quadrupole of each statistic. Our results show that each of the four marked power spectra can tighten the standard power spectrum constraints on the cosmological parameters $\Omega_{\rm m}$, $\Omega_{\rm b}$, $h$, $n_s$, $M_\nu$ by $15-25\%$ and on $\sigma_8$ by a factor of 2. The same analysis performed by combining the standard and four marked power spectra shows a substantial improvement compared to the power spectrum constraints that is equal to a factor of 6 for $\sigma_8$ and $2.5-3$ for the other parameters. Our constraints may be conservative, since the galaxy number density in the Molino catalogs is much lower than the ones in future galaxy surveys, which will allow them to probe lower density regions of the large-scale structure.

\end{abstract}



\section{Introduction} \label{sec:intro}

One of the main goals of current and future galaxy surveys is acquiring information about the matter content of our Universe, its accelerated expansion, its initial conditions, and the neutrino mass scale. Future surveys, such as Euclid\footnote{\url{https://www.euclid-ec.org}}, DESI\footnote{\url{https://www.desi.lbl.gov}}, and Roman Space Telescope\footnote{\url{https://roman.gsfc.nasa.gov/}}, will enable us to observe a very large fraction of the sky, up to redshift $z\sim 2-3$, giving us a very detailed map of the Universe. Large volumes coupled to high galaxy densities will allow us to measure the two-point statistics (correlation function and power spectrum) of galaxies with unprecedented accuracy, and will enable measurements of higher order point functions and different statistics. 

Our current understanding of the Universe \citep{Planck2018} is well described by the Standard Model of cosmology \citep{Dodelson_book}, which involves an early time where the distribution of matter fluctuations is very close to Gaussian, and later times where these fluctuations become non-Gaussian on small scales due to non-linear gravitational evolution, while remaining Gaussian on very large scales. A Gaussian field is entirely described by its two-point function, whereas this is not true
in general
for a non-Gaussian field. Galaxies trace the distribution of matter in a biased way and they are commonly used to infer the distribution of the underlying matter and the cosmological information that this distribution carries. Similarly to the matter field, the galaxy field is non-Gaussian at late times and on small scales. Therefore, standard analyses involving the galaxy correlation function, or power spectrum, cannot be exhaustive in retrieving cosmological and physical information from galaxy surveys. There are multiple ways to pick up information beyond the two-point statistics. 

A first method relies on measuring the higher-order point functions, such as bispectrum and trispectrum. This methodology can be computationally expensive---especially when modeling the covariance matrices, even though later achievements \citep{Philcox_Nfunc} showed that the analysis can be sped up---but is very promising \citep{Gilmarin_2017,Molino_Bk,Philcox_2020}. Another approach involves the use of different summary statistics, such as the number density of voids and clusters \citep{Pisani_2019,Bayer_2021,Kreisch_2021}, counts-in-cells (CIC) \citep{Bernardeau_1994,Sheth_1998,Uhlemann_2020}, which is the probability distribution function (PDF) of the mean matter density inside patches of a given size, Minimum Spanning Tree \citep{Naidoo_2020,Naidoo_2022}, scattering transforms \citep{Valogiannis_2021,Eickenberg_2022}, and many others. A different method considers separating regions of the Universe depending on their density, and computing the auto-correlation of regions with similar density (\cite{Repp_2022}) or cross-correlating density-selected regions with the total field (\cite{Paillas_2021}) --- a specific case being the void-galaxy cross-correlation, where regions devoid of galaxies are cross-correlated with the full galaxy field \citep[e.g.][]{Paz_2013,Hamaus_2015,Woodfinden_2022}. A further different approach concerns non-linear transformations of the galaxy field. Among these, the log-transformation aims to Gaussianize the field, such that most of the information is encoded in the two-point function of the log field \citep{Neyrinck_2009}. One other example of this type of transformations is the marked field, which is obtained by weighting an original field by a mark that depends on the local (smoothed) density around that point. The two-point statistic of the marked field, the so-called marked power spectrum, contains the two-point function as well as higher-order functions of the original field \citep{Philcox_2020}, since the mark is a non-linear function on the original density field. This last statistics, the marked power spectrum, is what we will investigate in this paper. 

Marked power spectra have been introduced in cosmology as desirable tools to study modifications of gravity \citep{White_2016,Valogiannis_2017,Armijo_2018,Hernandez-Aguayo_2018}. In a previous paper we highlighted the relevance of marks that up-weight low density regions to study cosmology in a broader sense \citep{Massara_2021}. In that work, we considered the matter density field, rather than the observable galaxy field, and found that particular marked power spectra can outperform the standard matter power spectrum in constraining all cosmological parameters, and in particular the neutrino mass scale. In this paper, we investigate marked power spectra in mock galaxy catalogs in redshift-space. We perform a Fisher analysis to quantify how well galaxy marked spectra can constrain the value of the cosmological parameters in comparison to the standard galaxy power spectrum. 

The paper is organized as follows. Section \ref{sec:Mk} defines the mark transformation and the marked power spectrum, while Section \ref{sec:sims} discusses the set of N-body simulations and galaxy catalogs used in this work. Section \ref{sec:fisher} explains the Fisher formalism used to quantify the information content in an observable, and Section \ref{sec:cosmo_constraints} shows the main results of the Fisher analysis. Section \ref{sec:discussion} discusses the results obtained and Section \ref{sec:conclusions} concludes the paper giving a summary of the results and highlighting future directions. 

\section{Marked power spectrum}
\label{sec:Mk}
Marked power spectra are two-point statistics of a transformed (marked) field, which is obtained by weighting the original field by a function $m$ (mark) that is position-dependent. These statistics have been introduced in \cite{Stoyan1984}, and have been applied to galaxy fields in the context of astrophysics and cosmology. On the astrophysical side, \cite{Beisbart_2000}, \cite{Sheth_2005}, \cite{Skibba_2005}, \cite{Balaguera-Antolinez:2013afa}, \cite{Sureshkumar_2021}, and \cite{Rutherford_2021} considered marks that depend on galaxy luminosity, morphology, type, color, spin, etc, to study the dependence of galaxy clustering on these properties, or to understand how they correlate with the local environment to enable tests on galaxy formation models. On the cosmological side, \cite{White_2016} introduced a mark that depends on the local density around each galaxy, rather than its own properties,
\begin{equation}
    m(\vec{x};R,p,\delta_s) = \left[ \frac{1+\delta_s}{1+\delta_s+\delta_R(\vec{x})}\right]^p \equiv \left[1+\frac{\delta_R(\vec{x})}{1+\delta_s}\right]^{-p}\, ,
    \label{eq:mark}
\end{equation}
where $\delta_R(\vec{x})$ is the local overdensity obtained by filtering the field with a Top-Hat window at scale $R$, $\delta_s$ is a parameter controlling how sensitive the mark is to the local overdensity, and $p$ is an exponent regulating how much the galaxies in a particular environmental density are weighted compared to the ones in other environments. 

This mark was first designed to detect modifications of gravity in galaxy surveys. Indeed, most modified-gravity models exhibit a screening mechanism that recover General Relativity (GR) on small scales/ high-density environments. Since the screening mechanisms are density-dependent and deviations from GR are expected to be larger in mild-to-low density regions, enhancing the cosmological signal coming from those environments seems to be a preferable and effective strategy to detect modifications of gravity. The density-dependent mark in Eq.~\ref{eq:mark} transforms the galaxy field allowing to up-weight different environments of the large-scale structure depending on the sign of the exponent $p$: $p>0$ ($<0$) up-weights low (high) density regions. The applicability and usefulness of a two-point function of the density-dependent marked field (marked correlation function) in the context of modified gravity have been studied in N-body simulations \citep{Valogiannis_2017,Armijo_2018,Hernandez-Aguayo_2018} and in the LOWZ galaxy catalog of the Sloan Digital Sky Survey III (SDSS III) Baryon Oscillation Spectroscopic Survey (BOSS) Data Release 12 (DR12) \citep{Satpathy_2019}.  

Recently, we proposed the usage of density-dependent marks to constrain cosmology and measure the neutrino mass scale \citep{Massara_2021}. Particle physics experiments can measure the difference between pairs of squared masses of the three neutrino species, and set the minimum value for the sum of the three neutrino masses to be $M_\nu \leq 0.06$ eV for normal hierarchy (two light neutrinos and a heavy one) and $M_\nu \leq 0.1$ eV for the inverted hierarchy (a light neutrino and two heavy ones) \citep{Lesgourgues:2018ncw}. Cosmological observations are expected to provide complementary measurements, since the growth of cosmic structure is affected by the presence of neutrinos and is sensitive to the sum of the neutrino masses. The neutrino-induced effect on the large-scale structure is expected to be the largest in low-density regions, where the amount of cold dark matter is small and neutrinos can represent (depending on their mass) a large fraction of the total matter. Conversely, high-density regions contain a large amount of cold dark matter; neutrinos, which are almost evenly distributed in the Universe, are a very small fraction of the total matter in them. This indicates that density-dependent marked fields can be desirable to extract information on the neutrino mass scale from the full matter or galaxy field. Moreover, the two-point statistic (correlation function or power spectrum) of the marked field contains higher order point functions of the original field, because of the presence of the local density $\delta_R(\vec{x})$ in the mark. Since the galaxy and matter fields are non-Gaussian at late times and small scales, higher-order statistics contain additional information on top of the two-point function, bringing better constraints on cosmological parameters and the neutrino masses. Another advantage of marked statistics is the possibility to compute different marked fields from the same initial one. Each marked field, having different mark parameters, will highlight different parts of the large-scale structure. The related marked power spectra will contain slightly different information, and possibly present different parameter degeneracies. Thus, combinations of multiple marked power spectra of the same initial field can break degeneracies and allow 
retrieving even more cosmological information.  

In \cite{Massara_2021} we quantified the information content present in marked power spectra of the matter density field by performing a Fisher analysis. We considered statistics of the cold dark matter field (CDM) and of the cold dark matter plus neutrino (total matter) fields, including all scales up to $k_{\rm max}=0.5~h{\rm Mpc}^{-1}$ in an analysis performed in N-body simulations with box size equal to $1~h^{-1} {\rm Gpc}$. Our results showed that the marked power spectrum with mark parameters $R=10~h^{-1}$Mpc, $p=2$, and $\delta_s = 0.25$ can improve the power spectrum constraints on all cosmological parameters by $2-3$ times when considering the cold dark matter field, and $4-10$ times when considering the total matter field. Moreover, combinations of the marked and standard power spectra or of multiple marked spectra bring even tighter constraints. The best constraint on the neutrino masses can be obtained with two marked spectra, and it yields an error on $M_\nu$ equal to $\sigma(M_\nu)=0.35$ eV with the cold dark matter field, and $\sigma(M_\nu)=0.01$ eV with the total matter field. The latter can be used to achieve a $\sim6\sigma$ detection of the minimum sum of the neutrino masses. Unfortunately, the cold dark matter and total matter field cannot be directly observed, but the galaxy field traces the cold dark matter distribution. Therefore, we may expect that marked spectra of cold dark matter and of galaxies contain similar information. The purpose of this work is to quantify the latter.

\section{Simulations}
\label{sec:sims}

We perform the analysis using the Molino \citep{Molino_Bk} galaxy catalogs, obtained by populating a subset of the Quijote N-body simulations \citep{Quijote} with galaxies via a Halo Occupation Distribution (HOD) framework. In this Section we describe the simulations and galaxy catalogs used for the analysis.

\subsection{N-body simulations}
Quijote are a set of more than $44,000$ N-body simulations run with many different cosmological models and constructed to perform Fisher analysis and to train machine learning models. The simulations are run using the TreePM code Gadget-III, a later version of the publicly available Gadget-II code \citep{Springel_2005}, starting from initial conditions (ICs) set at redshift $z=127$. Second-order Lagrangian perturbation theory (2LPT) is used to generate the ICs in all cosmologies, except for the massive neutrino ones where the Zel'dovich approximation is used. Additionally, a subset of simulations in the so-called fiducial cosmology have Zel'dovich ICs to reliably compute the numerical derivatives with respect to neutrino mass (for more details see Section~\ref{sec:derivatives}). 

Each simulation has a box size with length $1h^{-1}$Gpc and contains $512^3$ cold dark matter particles in the fiducial resolution setup, plus $512^3$ neutrino particles for cosmologies with massive neutrinos. The simulations can be divided into two main groups: (1) latin hypercubes and (2) sets of simulations where only one parameter is changed with respect to the fiducial cosmology that is set to be a flat $\Lambda$CDM cosmology with matter density parameter $\Omega_{\rm m}=0.3175$, baryon density parameter $\Omega_{\rm b}=0.049$, dimensionless Hubble constant $h=0.6711$, spectral index $n_s=0.9624$, linear matter fluctuation amplitude $\sigma_8=0.834$, and sum of neutrino masses $M_\nu=0$ eV. We will use a subset of the simulations in group (2) to perform our analysis, whose features are displayed in Table ~\ref{tab:quijote}. The $8,000$ boxes in fiducial cosmology will be used to compute covariance matrices (see Section~\ref{sec:covariance}), while the $500$ realizations in cosmologies with the variation of one cosmological parameter above or below its fiducial value will be used to compute numerical derivatives of the observables (standard or marked power spectra) with respect to each cosmological parameter (see Section~\ref{sec:derivatives}). 

In Quijote, halos are identified in the cold dark matter distribution using a Friends-of-Friends (FoF) algorithm \citep{Davis_1985} with linking length parameter $b = 0.2$. Only halos that contain at least 20 CDM particles are considered. This means that the minimum halo mass depends on the value of $\Omega_m$ and it is equal to $1.31\times10^{13}~h^{-1}$M$_\odot$ in the fiducial cosmology. 

\begin{table}[]
    \centering
    \newcolumntype{M}[1]{>{\centering\arraybackslash}m{#1}}
    \begin{tabular}{|M{2cm}|M{1cm}|M{1cm}|M{1cm}|M{1cm}|M{1cm}|M{1cm}|M{2cm}|M{2cm}|}
        \hline 
        Name &  $\Omega_m$ & $\Omega_b$ & $h$ & $n_s$ & $\sigma_8$ & $M_\nu$ &
        realizations & ICs \\
         &   &  &  &  &  & [eV] & &  \\
         \hline \hline
        Fiducial        & 0.3175 & 0.049 & 0.6711 & 0.9624 & 0.834 & 0 & 8,000 & 2LPT \\
        Fiducial ZA     & 0.3175 & 0.049 & 0.6711 & 0.9624 & 0.834 & 0 & 500 & Zel'dovich \\
                        \hline
        $\Omega_m^{+}$ & {\bf 0.3275} & 0.049 & 0.6711 & 0.9624 & 0.834 & 0 & 500 & 2LPT \\
        $\Omega_m^{-}$ & {\bf 0.3075} & 0.049 & 0.6711 & 0.9624 & 0.834 & 0 & 500 & 2LPT \\
        \hline
        $\Omega_b^{++}$ & 0.3175 & {\bf 0.051} & 0.6711 & 0.9624 & 0.834 & 0 & 500 & 2LPT \\
        $\Omega_b^{--}$ & 0.3175 & {\bf 0.047} & 0.6711 & 0.9624 & 0.834 & 0 & 500 & 2LPT \\
        \hline
        $h^{+}$         & 0.3175 & 0.049 & {\bf 0.6911} & 0.9624 & 0.834 & 0 & 500 & 2LPT \\
        $h^{-}$         & 0.3175 & 0.049 & {\bf 0.6511} & 0.9624 & 0.834 & 0 & 500 & 2LPT \\
        \hline
        $n_s^{+}$       & 0.3175 & 0.049 & 0.6711 & {\bf 0.9824} & 0.834 & 0 & 500 & 2LPT \\
        $n_s^{-}$       & 0.3175 & 0.049 & 0.6711 & {\bf 0.9424} & 0.834 & 0 & 500 & 2LPT \\
        \hline
        $\sigma_8^{+}$  & 0.3175 & 0.049 & 0.6711 & 0.9624 & {\bf 0.849} & 0 & 500 & 2LPT \\
        $\sigma_8^{-}$  & 0.3175 & 0.049 & 0.6711 & 0.9624 & {\bf 0.819} & 0 & 500 & 2LPT \\
        \hline
        $M_\nu^{+}$     & 0.3175 & 0.049 & 0.6711 & 0.9624 & 0.834 & {\bf 0.1} & 500 & Zel'dovich \\
        $M_\nu^{++}$    & 0.3175 & 0.049 & 0.6711 & 0.9624 & 0.834 & {\bf 0.2} & 500 & Zel'dovich \\
        $M_\nu^{+++}$   & 0.3175 & 0.049 & 0.6711 & 0.9624 & 0.834 & {\bf 0.4} & 500 & Zel'dovich \\
        \hline
    \end{tabular}
    \caption{Description of the N-body simulations used in the Fisher analysis. $\Omega_{\rm m}$ is the matter density parameter,  $\Omega_{\rm b}$ is the baryon density parameter, $h$ is the dimensionless Hubble constant, $n_s$ is the spectral index, $\sigma_8$ is the root-mean-square amplitude of the linear matter fluctuations at $8~h^{-1}$Mpc, and $M_\nu$ is the sum of neutrino masses.}
    \label{tab:quijote}
\end{table}

\subsection{Galaxy catalogs}

The Molino galaxy catalogs \citep{Molino_Bk} are built upon the Quijote simulations described above using an HOD framework that populates halos with galaxies. In this framework, the probability of a given halo to host $N$ galaxies depends only on its mass, $M_h$. Moreover, the sample is divided into central (placed at the center of a halo) and satellite (placed inside the halo to follow a NFW profile \citep{NFW}) galaxies. The specific HOD model implemented in Molino depends on 5 parameters ($\log M_{\rm min}, \sigma_{\log M} , \log M_0, \alpha, \log M_1$) as in \cite{Zheng_2007}. Only one central galaxy can be placed in each halo, thus centrals are drawn from a Bernoulli distribution with mean number $\langle N_c \rangle(M_h)$,
\begin{equation}
    \langle N_c \rangle = \frac{1}{2}\left[ 1- {\rm erf} \left(\frac{ \log M_h - \log M_{\rm min} }{\sigma_{\log M}}\right) \right] \, ,
\end{equation}
while the number of satellite galaxies follows a Poisson distribution with mean number $\langle N_s \rangle(M_h)$,
\begin{equation}
    \langle N_s \rangle = \langle N_c \rangle \left( \frac{M_h-M_0}{M_1}  \right)^\alpha\, .
\end{equation}
The parameter $M_{\rm min}$ indicates the halo mass scale where $\langle N_c \rangle$ goes from 0 to 1, and the width of this transition is described by $\sigma_{\log M}$. The mass $M_0$ corresponds to the minimum halo mass hosting a satellite galaxy and $M_h = M_0+ M_1$ is the typical halo mass with $1$ satellite galaxy.

The fiducial HOD parameters of the Molino catalogs at redshift $z=0$ are 
\begin{equation}
\left\{\log M_{\rm min}, \sigma_{\log M} , \log M_0, \alpha, \log M_1\right\} = \left\{13.65, \, 0.2, \, 14.0, \, 1.1, \, 14.0\right\}\, .
\end{equation}
These values have been chosen to mimic the best-fit HOD parameters to SDSS data from \cite{Zheng_2007}, but have been adapted to the low minimum halo mass limit of the Quijote suite. The fiducial HOD model has been applied to all the N-body simulations listed in Table~\ref{tab:quijote}. In order to compute derivatives with respect to HOD parameters, HOD models with variation of a single parameter above and below the fiducial value have been applied to 500 boxes in the fiducial cosmology. The step size of these variations are
\begin{equation}
    \left\{\Delta\log M_{\rm min}, \Delta\sigma_{\log M} , \Delta\log M_0, \Delta\alpha, \Delta\log M_1\right\} = \left\{0.05, \, 0.2, \, 0.2, \, 0.2, \, 0.2\right\}\, ,
\end{equation}
which have been tested to give convergent derivatives for the power spectrum and the bispectrum \citep{Molino_Bk}. We have considered additional step sizes and confirmed that the ones implemented in Molino are reliable also for the marked power spectra investigated in this paper. 

To compute derivatives, each simulation box is populated with galaxies in 5 different ways (initial seeds) and the redshift-space standard and marked power spectra are measured using 3 different lines-of-sight per catalog. Thus, each simulation gives rise to 5 galaxy catalogs and 15 measurements of a given statistics, allowing us to obtain a more accurate estimation of its derivatives. The same procedure cannot be implemented to compute covariances, since each data vector must be measured from independent realizations. For covariances, we will use as many data vectors as the number of N-body simulations available in the fiducial cosmology set (8,000).

\section{Fisher Formalism}
\label{sec:fisher}

We use the Fisher information matrix to describe the information content of different statistics. For statistics that follow a multivariate Gaussian distribution, the Fisher matrix can be computed as
\begin{equation}
    F_{ij} = \Sigma_{\alpha,\beta} \, \frac{\partial S_\alpha}{\partial \theta_i} \, C^{-1}_{\alpha \beta}\, \frac{\partial S_\beta}{\partial \theta_j}
    \label{eq:fisher}
\end{equation}
where $\vec{S}=\left\{S_0,S_1,... \right\}$ is the data vector that can be a single statistic or the concatenation of many of them (standard and/or marked power spectrum) measured at different wavenumbers $k$, $\vec{\theta}$ is the collection of cosmological and HOD parameters, and $C$ is the covariance matrix (see Section~\ref{sec:covariance}). The partial derivatives of each data vector are discussed in Section~\ref{sec:derivatives}. The Fisher matrix gives the variance of an optimal unbiased estimator for the parameter $\theta_i$: $\sigma^2(\theta_i) \geqslant (F^{-1})_{ii}$. The lower the marginalised error, the better a certain parameter can be constrained by the considered statistic(s). For each statistic we calculate the monopole and quadrupole up to a maximum wavenumber equal to $k_{\rm max}=0.5~h~ {\rm Mpc}^{-1}$, and we subtract the Poisson shot-noise $1/\bar{n}_g$, where ${n}_g$ is the galaxy number density, to the monopole of the power spectrum computed from each mock catalog. We do not subtract any estimation of the shot-noise to the marked power spectra, since its linear prediction  \citep[see][]{Philcox_2020,Philcox_2021} is not accurate on the considered scales. We test the impact of shot-noise on the Fisher constraints obtained with marked power spectra in Section~\ref{sec:appendix}, and find
no dependence on the number density of galaxies in the catalog. 

Different mark models allow us to perform different transformations of the galaxy field; the corresponding marked power spectra will contain different information about the cosmological parameters, and they will exhibit different parameter degeneracies. Here, we consider 60 mark models obtained by combining in 60 different ways the following values for the mark parameters: $R = 10,15,20,25,30 \,h^{-1}$Mpc, $p = 1,2,3,4$, $\delta_s= 0.1,0.25,0.5$. Thus, the monopole and quadrupole of the 60 marked spectra can be measured in each galaxy mock, and used to perform the Fisher analysis. We identify the marked spectra that allow to put stringent constraints on the cosmological parameters and show their parameter values in Table \ref{tab:best_Mk}. In Section~\ref{sec:cosmo_constraints} we will discuss their constraining power in comparison to the standard galaxy power spectrum. Note that some other mark models with $p=4$ give better constraints but they exhibit noisy derivatives and covariance matrices with strong non-diagonal elements; we decided to discard these models to make sure that the final results are reliable. 

For all of the considered values of the smoothing scale $R$, the corresponding wavenumber $2\pi/R$ is smaller than the maximum wavenumber considered in the analysis: $k_{\rm max} = 0.5~h~{\rm Mpc}^{-1}$. A $k$-space Top-Hat filter would correctly restrict the range of wavenumbers to exclude the contributions from high-$k$ modes in the analysis, but that filter is not localized in configuration space. Nonetheless, considering a Top-Hat in configuration space with large $R$ values prevents the inclusion of most of the information coming from $k>0.5~h~{\rm Mpc}^{-1}$ to the analysis. In principle, small smoothing scales would allow one to obtain a more detailed estimation of the local density, while very large values of $R$ correspond to smoothing overdensities close to zero. However, the values of $R$ cannot be arbitrary small not only because of the need to control the maximum wavenumber included in the analysis, but also because of limitations of the N-body simulations (the grid in the initial condition has size $2~h^{-1}$Mpc), and the sparsity of galaxies: Very small $R$ translates into an estimation of $\delta_R$ that may not reflect the environmental density, but just be a Poisson noise realization of it. Smoothing scales $R\geq20 ~h^{-1}$Mpc ensure that every sphere has on average more than $5$ galaxies and it is well above the grid size of the ICs. 

\begin{table}[]
    \centering
    \newcolumntype{M}[1]{>{\centering\arraybackslash}m{#1}}
    \begin{tabular}{|M{2cm}|M{2cm}|M{2cm}|M{2cm}|M{2cm}|}
        \hline 
        M(k) &  R[$h^{-1}$Mpc] & p & $\delta_s $ \\
         \hline \hline
        $M_1$ & 30 & 1 & 0.1 \\
        $M_2$ & 25 & 1 & 0.25 \\
        $M_3$ & 20 & 1 & 0.5\\
        $M_4$ & 30 & 1 & 0.5 \\
         \hline
    \end{tabular}
    \caption{Values for the mark parameters ($R, p, \delta_s$) of selected marked power spectra $M_{1}$, $M_2$, $M_3$, and $M_4$.}
    \label{tab:best_Mk}
\end{table}

\subsection{Covariance}
\label{sec:covariance}

The covariance matrix $C$ of different statistics is measured as
\begin{equation}
    C_{\alpha \beta} = \langle \, \left[ S_\alpha - \bar{S}_\alpha\right]\left[S_\beta - \bar{S}_\beta\right]\, \rangle
\end{equation}
where $\langle...\rangle$ indicates the average over different realizations, $\bar{S}_i = \langle S_i\rangle$, and $\vec{S}$ is the data vector containing one or multiple concatenated statistics evaluated at various wavenumbers $k$. We
use 8,000 galaxy mock catalogs built upon the N-body simulations in fiducial cosmology to compute the covariance. Convergence tests
of the covariance matrix are discussed in Appendix~\ref{sec:appendix}.

Figure~\ref{fig:cov} shows the correlation matrix ($C_{\alpha\beta}/\sqrt{C_{\alpha\alpha}C_{\beta\beta}}$) for a data vector $\vec{S} = \left\{P_0,P_2,M_{1,0},M_{1,2},..,M_{4,0},M_{4,2}\right\}$ obtained by concatenating the monopole and quadrupole of the power spectrum with the multipoles of the four marked spectra selected in Table~\ref{tab:fisher}. All statistics are considered up to a maximum wavenumber $k_{\rm max} = 0.5~h$ Mpc$^{-1}$. The four bottom left blocks display the correlation matrix of the power spectrum, showing that the auto-covariance of both the monopole and quadrupole are not diagonal, and their cross-covariance is anticorrelated. Moving towards the right (or the upper part of the plot), the plot displays the blocks corresponding to the marked power spectra and their cross-correlation with the power spectrum. We find that the auto-correlation matrix of the monopole of $M_1$ exhibits positive off-diagonal elements on small scales, while the auto-correlation matrix of its quadrupole is almost diagonal. The marked power spectra $M_2$, $M_3$ and $M_4$ have auto-correlation matrices that are also almost diagonal. All the blocks quantifying the cross-correlations between monopole and quadrupoles of different statistics are mostly diagonal, with the monopole-monopole and quadrupole-quadrupole off-diagonal components being slightly correlated and the monopole-quadrupole parts being
uncorrelated
or slightly anti-correlated.

\begin{figure}[ht!]
\centering
\includegraphics[width=0.65\textwidth]{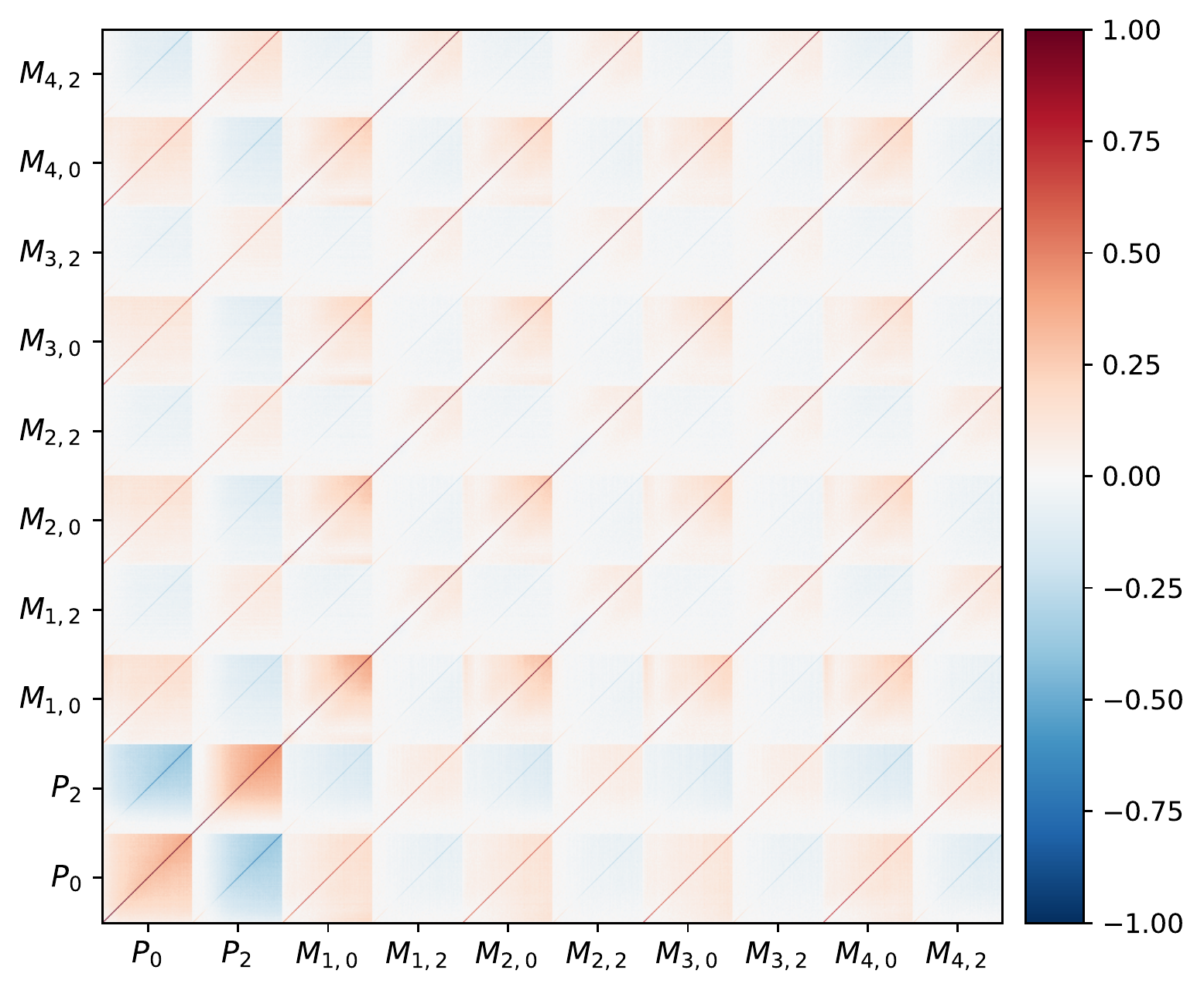}
\caption{Correlation matrix of the monopole and quadrupole of the power spectrum and four marked power spectra $M_i$ with the following parameters. $M_1$: $R=30~h^{-1}$Mpc, $p=1$, and $\delta_s=0.1$. $M_2$: $R=25~h^{-1}$Mpc, $p=1$, and $\delta_s=0.25$. $M_3$: $R=20~h^{-1}$Mpc, $p=1$, and $\delta_s=0.5$. $M_4$: $R=30~h^{-1}$Mpc, $p=1$, and $\delta_s=0.5$. All observables are considered up to $k_{\rm max}= 0.5~h~{\rm Mpc}^{-1}$. \label{fig:cov}}
\end{figure}

\subsection{Derivatives}
\label{sec:derivatives}

\begin{figure}[ht!]
\gridline{\fig{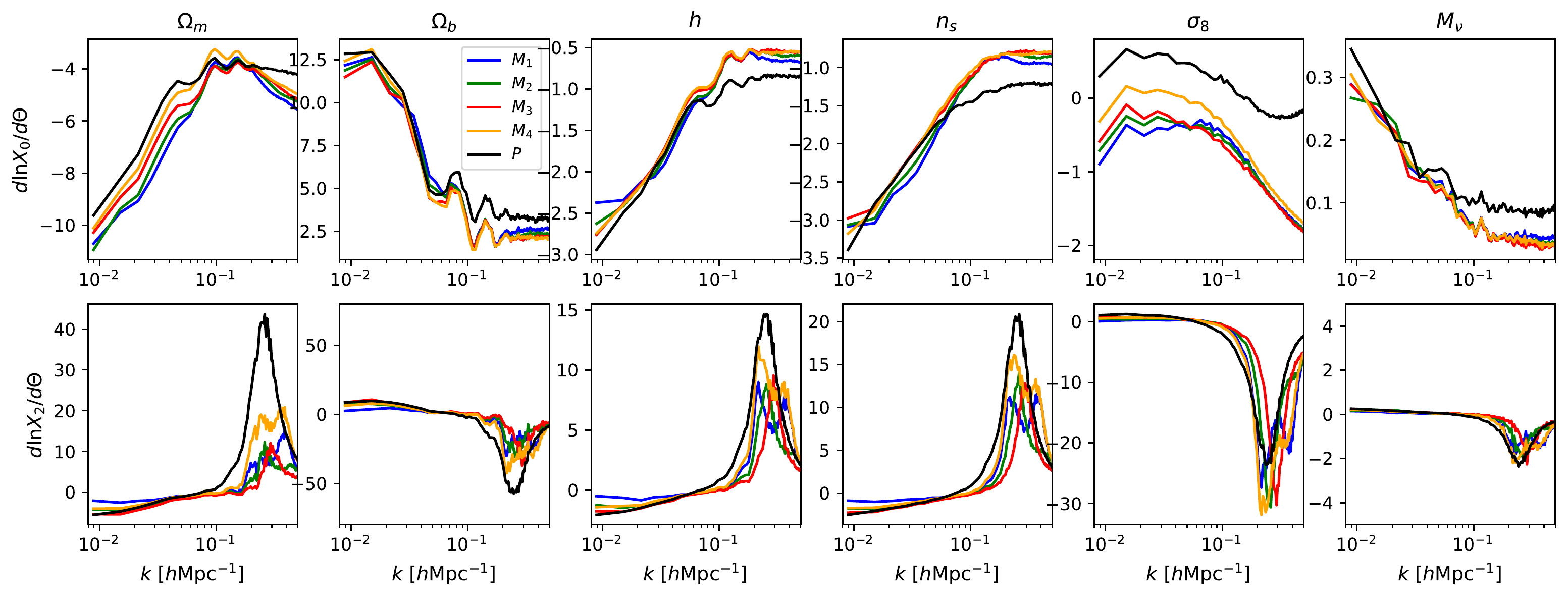}{0.99\textwidth}{(a)}
         }
\gridline{\fig{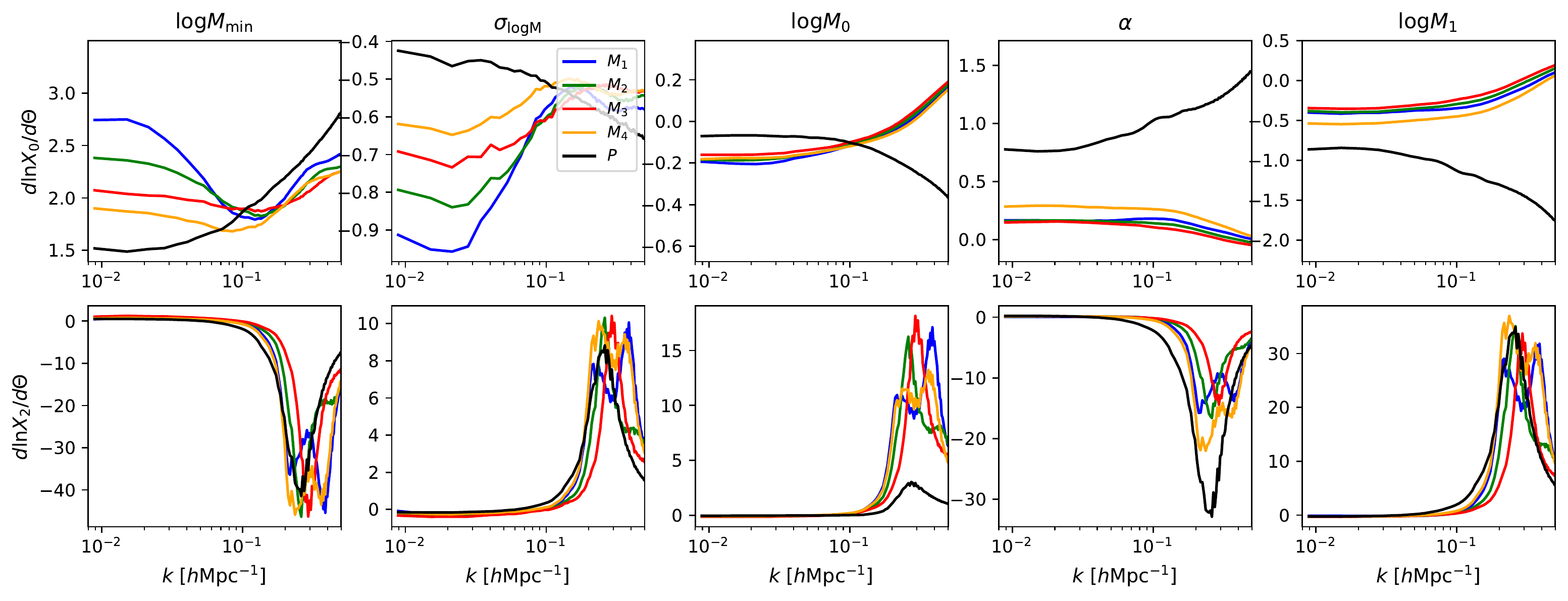}{0.99\textwidth}{(b)}
         } 
\caption{(a) Numerical logarithmic derivatives with respect to cosmological parameters of the monopole (top) and quadrupole (bottom) of power spectrum (black) and four marked power spectra $M_i$ (colored) described in Table~\ref{tab:best_Mk}. (b) Numerical logarithmic derivatives for the same statistics with respect to HOD parameters. \label{fig:deriv} }
\end{figure}

We compute the partial derivatives of Equation~\ref{eq:fisher} numerically via
\begin{equation}
    \frac{\partial \vec{S}}{\partial \theta} \simeq \frac{\vec{S}(\theta+\delta\theta)-\vec{S}(\theta-\delta\theta)}{2\, \delta\theta}\, ,
\end{equation}
using the simulations and HOD realizations where the considered parameter $\theta$ assumes values above and below the fiducial one.
This can be done for all parameters, except for the neutrino masses, since its fiducial value is $M_\nu = 0$ and negative masses do not have any physical meaning. In order to compute these derivatives, we use statistics computed in cosmologies with different positive values for the neutrino masses: $M_\nu^{+} = 0.1$ eV, $M_\nu^{++} = 0.2$ eV, and $M_\nu^{+++} = 0.4$ eV. We can use different estimators using these values: 
\begin{eqnarray}
\label{eq:deriv_Mnu}
\left[\frac{\partial \vec{S}}{\partial M_\nu}\right]_1 &\simeq& \frac{\vec{S}(dM_\nu)-\vec{S}({M}_\nu=0)}{dM_\nu} + \mathcal{O}(dM)\\\nonumber
\left[\frac{\partial \vec{S}}{\partial M_\nu}\right]_2 &\simeq& \frac{-2\vec{S}(2dM_\nu)+4\vec{S}(dM_\nu)-3\vec{S}({M}_\nu=0)}{2dM_\nu} + \mathcal{O}(dM^2)\\\nonumber
\left[\frac{\partial \vec{S}}{\partial M_\nu}\right]_3 &\simeq&
\frac{\vec{S}(4dM_\nu)-12\vec{S}(2dM_\nu)+32\vec{S}(dM_\nu)-21\vec{S}({M}_\nu=0)}{12dM_\nu} + \mathcal{O}(dM^3)
\end{eqnarray}
where $dM_\nu$ takes values $0.1,0.2,0.4$ eV in the first equation, $0.1,0.2$ eV in the second equation, and $0.1$ eV in the last one. All these estimators have different levels of error: in the case of a noiseless mean data vectors $\vec{S}$, they increase with $dM_\nu$ and decrease with the number associated to the estimator. The data vectors are not noiseless; we assume the definition number 3 in the main analysis and test the stability of our results when changing the estimator in Section~\ref{sec:appendix}. As mentioned in Section~\ref{sec:sims}, the simulations run in massive neutrino cosmologies have Zel'dovich initial conditions, while the others have 2LPT ICs. In Equation~\ref{eq:deriv_Mnu} the statistics must be measured from both massive and massless neutrino cosmologies. To compute these derivatives we use the simulations in fiducial cosmology but with Zel'dovich ICs --- otherwise we may
introduce a signal in the derivatives that comes from using simulations with different methods to generate the initial conditions \citep[see][for further details]{Quijote}. 

The derivatives with respect to all parameters are computed as the mean of 7,500 measurements, since that is the number of different redshift-space galaxy mock catalogs available in each cosmology/HOD model (note that they are not independent realizations). Figure~\ref{fig:deriv} displays the logarithmic derivatives of the power spectrum (black) and of the four marked spectra (colored) with respect to the cosmological parameters (plot a) and the HOD parameters (plot b). The top and bottom panels in each plot show the derivatives of the monopoles and quadrupoles, respectively. The logarithmic derivatives of all the monopoles with respect to cosmology present similar but shifted features, except for the case of $\sigma_8$ and $\Omega_m$ where on small scales the standard and marked power spectra derivatives assume a different shape. On the other hand, derivatives of the monopole of marked (colored) and standard (black) spectra with respect to HOD parameters are very different, and differences among marked models are more pronounced for parameters controlling the central galaxies, $\log M_{\rm min}$ and $\sigma_{\log M}$. The logarithmic derivatives of the quadrupoles present spike features on small scales; the number and position of these spikes differ among the various statistics indicating that they contain different information about the galaxy bias scheme and cosmology. Even if the logarithmic derivatives contain clear features, they also present a non-negligible level of noise. We test the convergence of these derivatives in Section~\ref{sec:appendix} by testing the stability of the Fisher analysis results when changing the number of mocks used to compute the derivatives. Even if convergence seems to be reached, it could not be exact or at the desirable level of convergence. In that case, our results for the cosmological constraints could be optimistic. 

\section{Cosmological constraints}
\label{sec:cosmo_constraints}

\begin{figure}[ht!]
\centering
\includegraphics[width=0.9\textwidth]{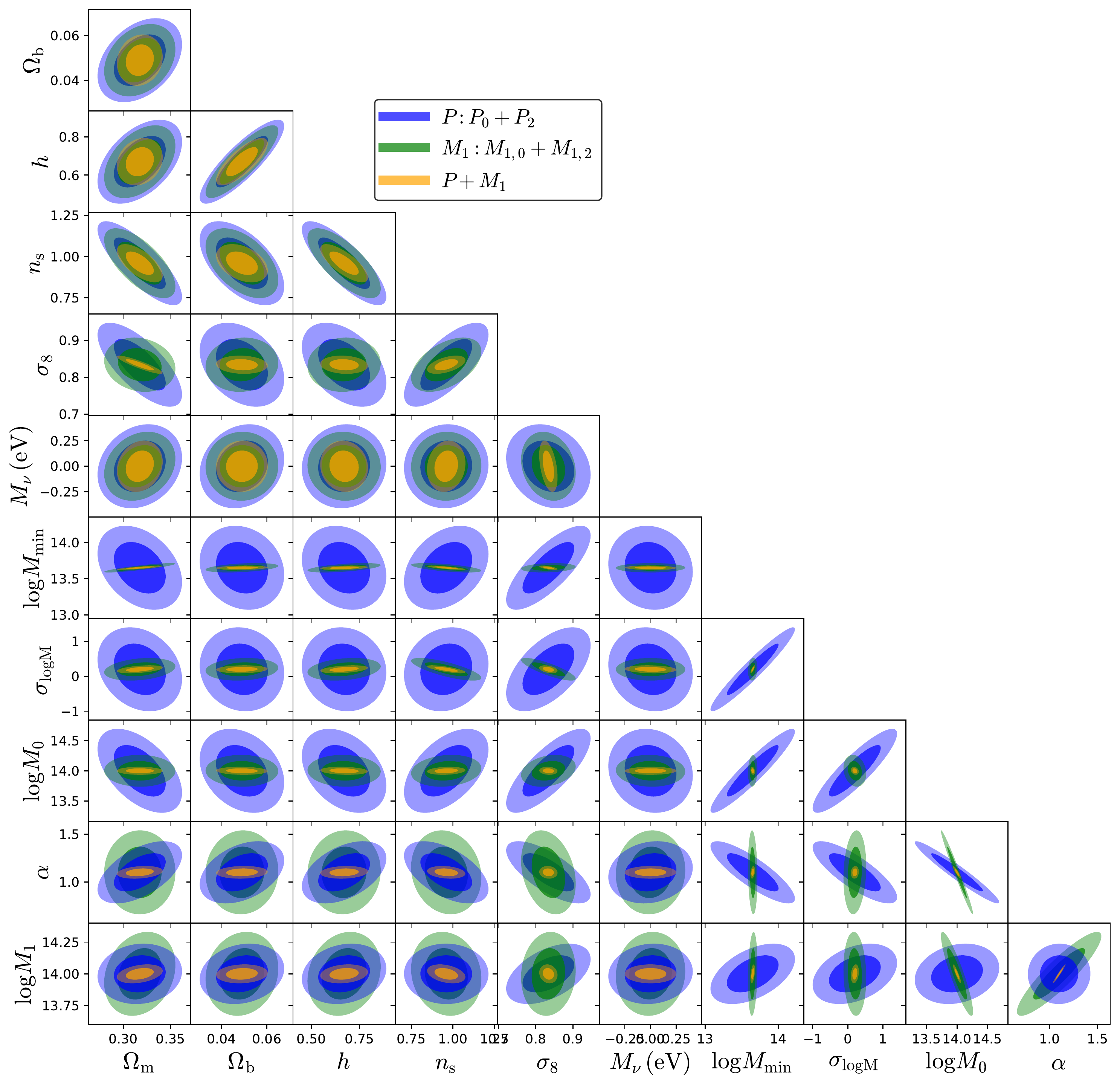}
\caption{Fisher matrix constraints on cosmological and HOD parameters obtained with galaxy power spectrum (blue), galaxy marked spectrum (green), and their combination (yellow). Darker and lighter shades are the $68\%$ and $95\%$ confidence contours. The mark model considered here ($M_1$) has mark parameters: $R=30~h^{-1}$Mpc, $p=1$, and $\delta_s=0.1$. The maximum wavenumber included for both monopole and quadrupole of each observable is $k_{\rm max} = 0.5~h~{Mpc}^{-1}$. \label{fig:fisher_PM}}
\end{figure}

We present here the results of the Fisher analysis using the four mark models identified in Table~\ref{tab:best_Mk}, and compare these finding to the Fisher forecast for the standard power spectrum. 

Figure \ref{fig:fisher_PM} shows the $1\sigma$ and $2\sigma$ constraints on cosmological and HOD parameters obtained from the Fisher analysis of standard power spectrum (blue) and the marked power spectrum $M_1$ (green); their combination is shown in orange. Both the monopole and the quadrupole of each observable have been included up to a maximum wavenumber $k_{\rm max} = 0.5~h~{\rm Mpc}^{-1}$. The standard power spectrum and $M_1$ exhibit different parameter degeneracies, and have different constraining power: While the power spectrum better constrains $\log M_1$ and $\alpha$, the marked spectrum $M_1$ better constrains the other parameters. $M_1$ up-weights low density regions defined on a $30~h^{-1}$Mpc scale. In these regions we expect to have mainly central galaxies, since satellites live in high-mass halos which should mainly trace high-density regions. It is therefore expected that $M_1$ can accurately constrain the HOD parameters controlling the number of central galaxies ($\log M_{min}$ and $\sigma_{\log M}$), while it is less sensitive to the HOD parameters controlling the number of satellite galaxies. The opposite applies to the power spectrum, which is mostly probing high-density regions. Combining power spectrum and $M_1$ will allow to use the information coming from both probes and eventually break some of their specific parameter degeneracies, as the orange contours highlight. We obtained similar plots when comparing the results from the other marked models $M_2$, $M_3$, and $M_4$ to one obtained with the power spectrum.

\begin{table}
\centering
\begin{tabular}{|C|CC|CC|CC|CC|CC|C|C|C|}
\hline
$\Theta$ & \multicolumn2c{$P$} & \multicolumn2c{$M_1$} & \multicolumn2c{$M_2$} & \multicolumn2c{$M_3$} & \multicolumn2c{$M_4$}  & $\Sigma_i M_i$ & $P+\Sigma_i M_i$ & $P/(P+\Sigma_i M_i)$ \\ 
 & 0+2 & 0 & 0+ 2 & 0  & 0+ 2 & 0 & 0+ 2 & 0 & 0+ 2 & 0 & 0+2 & 0+2 & 0+2 \\
\hline\hline
\Omega_m  & 0.036 & 0.045 & 0.030 & 0.039      & 0.030      & 0.039 & 0.029       & 0.039 & 0.029 & 0.038  & 0.016 & 0.015 & 2.4  \\
\Omega_b  & 0.015 & 0.019 & 0.013 & 0.016     & 0.013      & 0.017 & 0.013 & 0.018 & 0.013 & 0.016 & 0.007 & 0.006 & 2.5  \\
h         & 0.18  & 0.23  & 0.15  & 0.20      & 0.15       & 0.21  & 0.15  & 0.21 & 0.15  & 0.20  & 0.07 & 0.07 & 2.6 \\
n_s       & 0.20  & 0.26  & 0.17  & 0.21      &  0.17 & 0.22  & 0.16        & 0.23 & 0.17  & 0.22  & 0.06 & 0.06 & 3.3  \\
\sigma_8  & 0.091 & 0.15  & 0.059 & 0.18      & 0.045      & 0.19  & 0.041 & 0.20 & 0.047 & 0.18  & 0.017 & 0.015 & 6.1  \\
M_\nu     & 0.33  & 0.50  & 0.27 & 0.34 & 0.28       & 0.35  & 0.28        & 0.35 & 0.29  & 0.38  & 0.12 & 0.11 & 3.0  \\
\hline
\end{tabular}
\caption{Marginalized errors of cosmological parameters from different observables, when including the modes with $k < k_{\rm max} = 0.5~h~{\rm Mpc}^{-1}$.}
\label{tab:fisher}
\end{table}

\begin{figure}[ht!]
\centering
\includegraphics[width=0.9\textwidth]{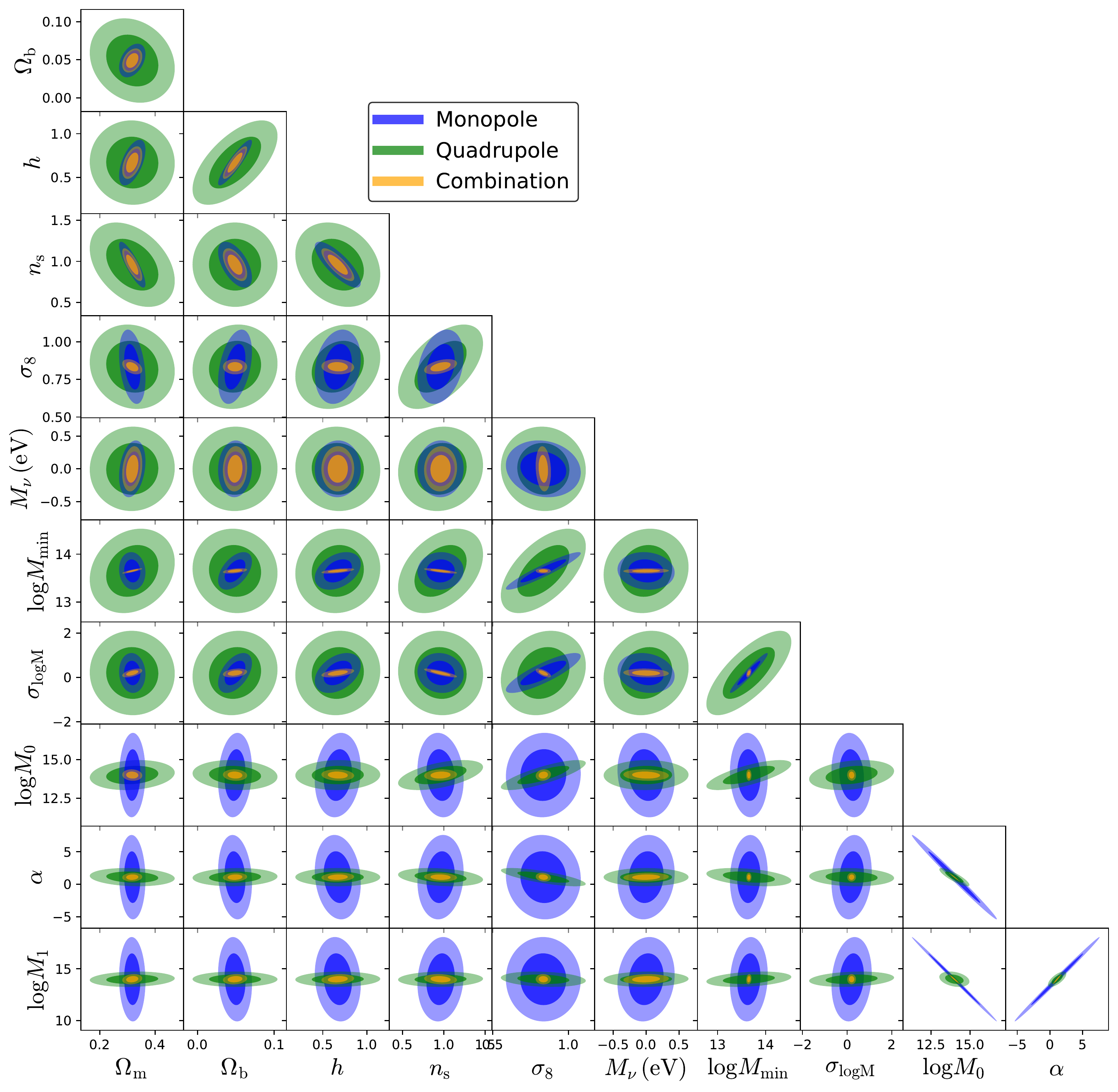}
\caption{Fisher matrix constraints (darker and lighter shades being the $68\%$ and $95\%$ confidence contours) on cosmological and HOD parameters obtained with the monopole (blue) and the quadrupole (green) of the marked power spectrum $M_3$: $R=20~h^{-1}$Mpc, $p=1$, and $\delta_s=0.5$. The combination of monopole and quadrupole is shown in yellow.\label{fig:fisher_M3-mono-quad}}
\end{figure}

\begin{figure}[h!]
\centering
\includegraphics[width=0.9\textwidth]{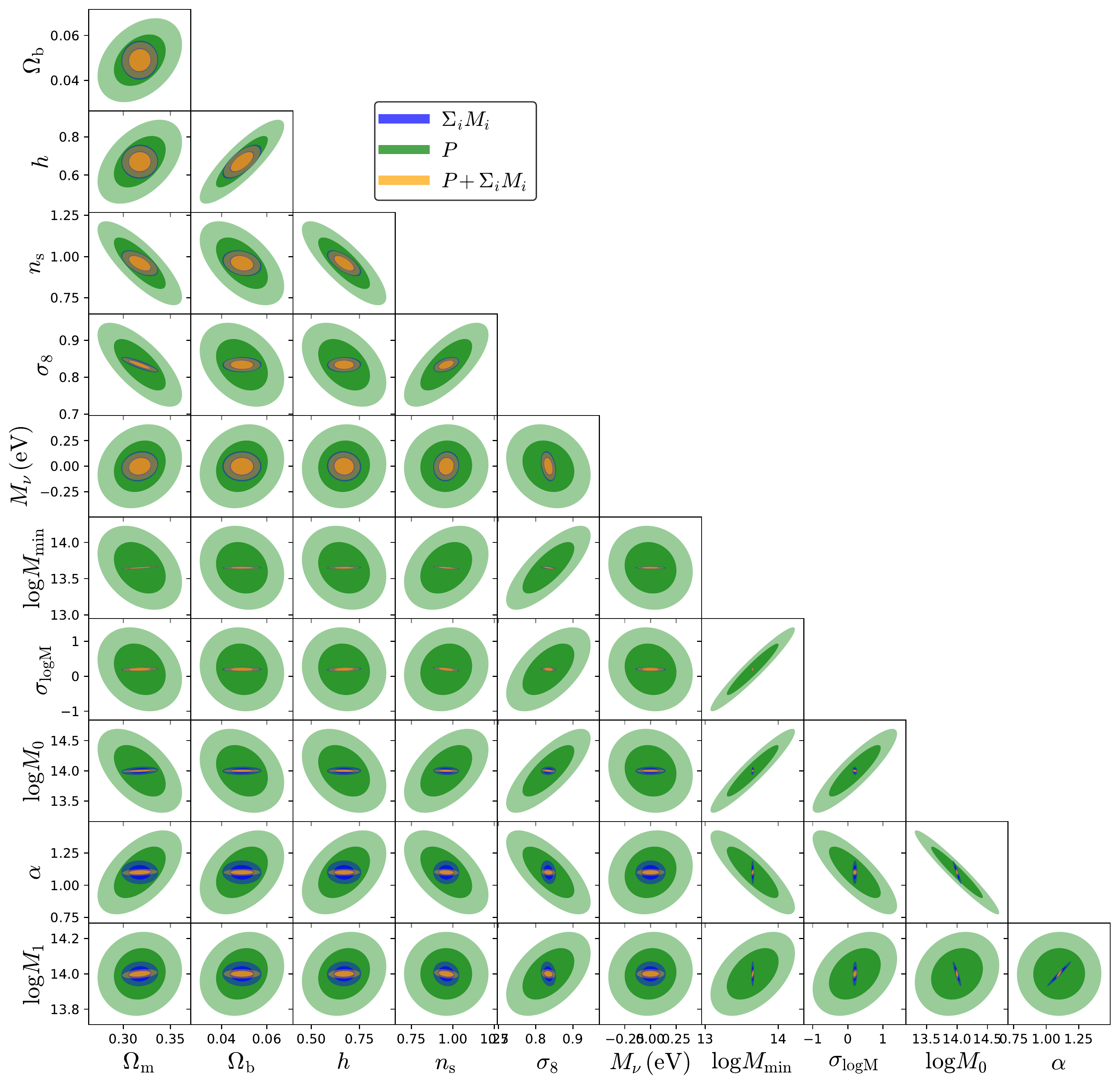}
\caption{Fisher matrix constraints (darker and lighter shades being the $68\%$ and $95\%$ confidence contours) on cosmological and HOD parameters obtained with the four marked power spectra $M_i$ in Table~\ref{tab:best_Mk} (blue), the power spectrum (green) and their combination (orange). \label{fig:fisher_4Mk}}
\end{figure}

\begin{figure}[h!]
\includegraphics[width=1\columnwidth]{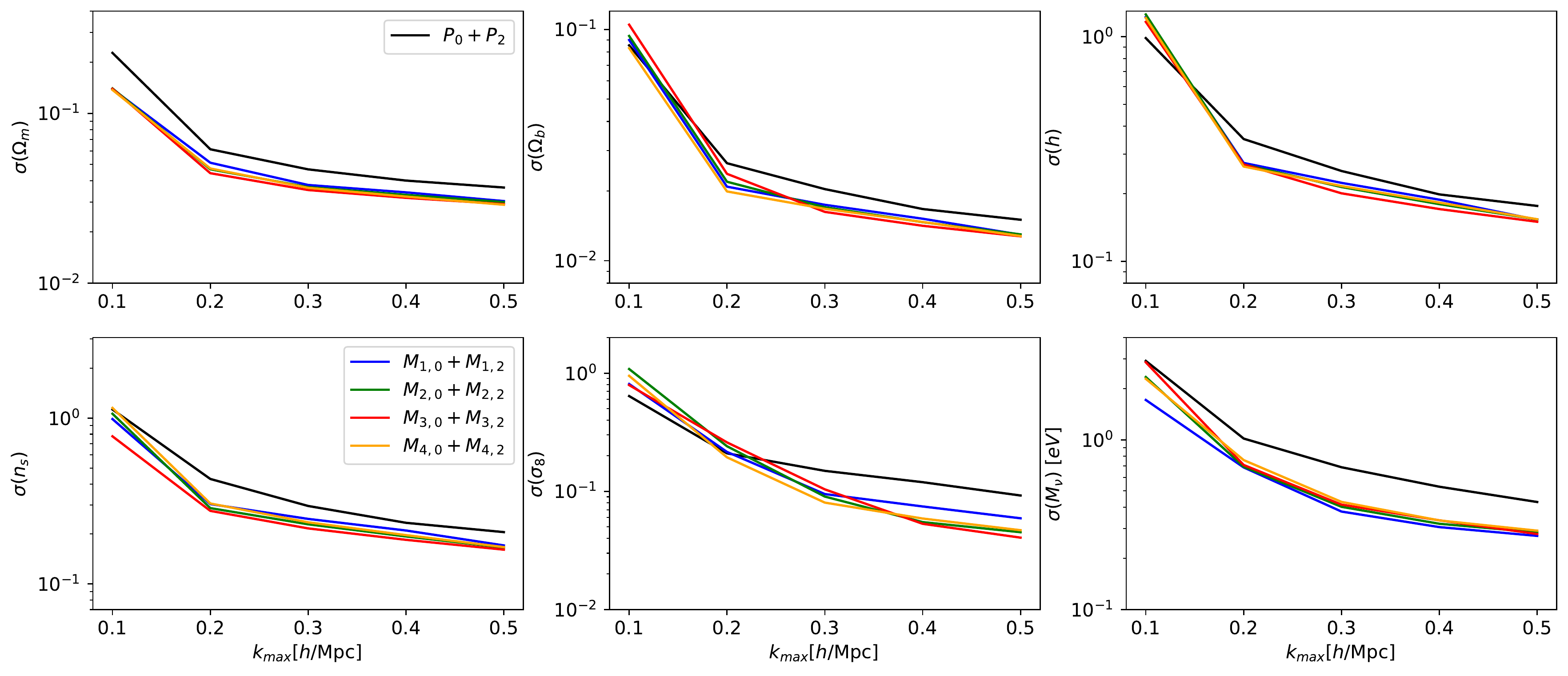}
\caption{Marginalized errors of cosmological parameters from different statistics as a function of the maximum wavenumber $k_{\rm max}$ included in the analysis. Results are computed using the monopole and quadrupole of each observable: power spectrum (black), marked power spectrum $M_1$ having $R=30~h^{-1}$Mpc, $p=1$, and $\delta_s=0.1$ (blue), marked power spectrum $M_2$ having $R=25~h^{-1}$Mpc, $p=1$, and $\delta_s=0.25$ (green), marked power spectrum $M_3$ having $R=20~h^{-1}$Mpc, $p=1$, and $\delta_s=0.5$ (red), and marked power spectrum $M_4$ having $R=30~h^{-1}$Mpc, $p=1$, and $\delta_s=0.5$ (orange).\label{fig:fisher_kmax}}
\end{figure}

Table \ref{tab:fisher} reports the marginalized errors on each cosmological parameter obtained from different observables up to wavenumbers $k<k_{\rm max}=0.5~h~{\rm Mpc}^{-1}$, and their combination. When considering a single observable, the table shows the errors obtained both from the monopole alone (right columns labeled as ``0") and from the combination of monopole and quadrupole (left columns labeled as ``$0+2$"). The quadrupole carries information about the anisotropic pattern of redshift space distortion generated by the velocity field. The galaxy motion depends on $\Omega_m$ and $M_\nu$ through the linear growth factor $f$, and on the amplitude of matter fluctuations $\sigma_8$. While the errors of most of the cosmological parameters do not decrease by more than $30-40\%$ when adding the quadrupole to the analysis, the errors on $\sigma_8$ exhibit a much larger improvement. In particular, when considering the power spectrum, the inclusion of the quadrupole improves the monopole error on $\Omega_m$ by $25\%$, on $\sigma_8$ by $65\%$, and on $M_\nu$ by $50\%$. The marked power spectra $M_i$ exhibit larger improvements on $\Omega_m$ ($30-35\%$) and specially on $\sigma_8$ (3 to 5 times smaller), while the improvement on $M_\nu$ is about $25-30\%$ but their monopole carries already more information than the monopole of the power spectrum. Figure \ref{fig:fisher_M3-mono-quad} shows the 2D constraints obtained with the monopole (blue) and quadrupole (green) of $M_3$. It is interesting to note that the quadrupole constrains the HOD parameters $\log M_0$, $\alpha$, and $\log M_1$ better than the monopole (but not as well as the multiples of the power spectrum). These parameters describe the number of satellite galaxies, which are responsible for the small scales redshift-space distortion appearing as Finger-of-God in the galaxy power spectrum. The monopole is instead more sensitive to the HOD parameters controlling the central galaxies, ($\log M_{min}$ and $\sigma_{\log M}$). The different degeneracy's directions present in the monopole and in the quadrupole in the planes of the HOD parameters allow the combination of these two probes to set tight constraints on all HOD parameters, and thus the galaxy bias. This translates into tight constraints on all cosmological parameters, and specially on $\sigma_8$.

The right columns of Table \ref{tab:fisher} display the results for the combination of four marked power spectra and of the standard plus the four marked spectra. The last column shows the ratio between constraints from the power spectrum and from the combination of all the statistics considered in this paper. Thus, it quantifies the improvement of their constraining power compared to the one of the power spectrum alone. The larger improvement is achieved for $\sigma_8$, where the errors shrink by a factor of 6.1, while the errors on the other cosmological parameters decrease by a factor of $2-3$. The different parameter degeneracies exhibited by the combination of the four marked power spectra and power spectrum with them is displayed in Figure~\ref{fig:fisher_4Mk} by the blue and orange contours, respectively. The 2D constraints obtained with the power spectrum alone are shown in green. We can note that the combination of marked power spectra is very powerful in setting constraints on all cosmological parameters and on the HOD parameters $\log M_{\rm min}$ and $\sigma_{\log M}$ controlling the number of central galaxies, while the addition of the power spectrum is important in improving the constraints on $\log M_0$, $\alpha$, and $\log M_1$ describing the number of satellite galaxies. It is also insightful to look at the dependence of the Fisher error on the maximum wavenumber considered, as shown in Figure \ref{fig:fisher_kmax}. The plotted marginalized errors have been obtained considering both the monopole and quadrupole of for power spectrum (black), and of the four marked spectra (colored). Including smaller wavenumbers increases the information and decreases the marginalized errors. $M_2$, $M_3$, and $M_4$ are particularly able to extract information on $\sigma_8$ from small scales, corresponding to $k>0.3~h~{\rm Mpc}^{-1}$. All the considered marked power spectra show similar information on the neutrino masses $M_\nu$, which is predominantly coming from scales $k>0.2~h~{\rm Mpc}^{-1}$.

\section{Discussion}
\label{sec:discussion}

\subsection{Comparison with constraints in the cold dark matter field}
This paper studies the constraining power of marked power spectra measured from the redshift space galaxy field of the Quijote simulations. In \cite{Massara_2021} we reported the results from an analogous Fisher analysis performed using the real space (no information on velocities) cold dark matter field of the same Quijote simulations. In that analysis, the tightest constraints on cosmological parameters had been obtained by combining two marked power spectra with parameters: $R=10~h^{-1}$Mpc, $p=2$, $\delta_s = 0.25$, and $R=10~h^{-1}$Mpc, $p=1$, and $\delta_s = 0$. The ratios between each marginalized cosmological errors obtained from the standard power spectrum and from the combination of the two marked spectra were: $3.3$ for $\Omega_m$, $2$ for $\Omega_b$, $2.4$ for $h$, $3.5$ for $n_s$, $3.8$ for $\sigma_8$ and $4$ for $M_\nu$. It is not straightforward to compare those results with the ones in Table \ref{tab:fisher} (last column), since the number of considered observables is different. However, we can note that the ratios (improvements with respect to the power spectrum constraints) obtained with the cold dark matter field analysis in \cite{Massara_2021} and with the galaxy field analysis in this paper are similar for all cosmological parameters, except for $\sigma_8$. In particular the galaxy field presents larger ratios than the cold dark matter field for $\Omega_b$, $h$, and $\sigma_8$, and smaller ratios for $\Omega_m$, $n_s$, and $M_\nu$.

Measuring statistics in redshift-space allows us to gather information from the velocity field, that depends on $f\sigma_8$ on linear scales. In the case of biased tracers, the amplitude of the density field depends on the combination of the linear galaxy bias and $\sigma_8$, while the velocity field still depends on the combination $f\sigma_8$. Therefore, redshift-space measurements of the galaxy power spectrum break the galaxy bias-$\sigma_8$ degeneracy and allow to place tighter constrain on $\sigma_8$. Marked power spectra, such as the $M_i$ considered here, exhibit the same degeneracy breaking, as discussed in Section \ref{sec:cosmo_constraints}, and they seem to extract much more information on $\sigma_8$ from the velocity field than the power spectrum. Indeed, while including the quadrupole decreases the monopole $\sigma_8$ error by a factor equal to $1.6$ when considering the power spectrum, it decreases the monopole errors by $3-5$ times when considering the marked power spectra $M_i$.

Let us consider $\Omega_m$ and $M_\nu$. The combination of the four selected galaxy marked spectra and the galaxy power spectrum ($P+\Sigma_i M_i$) allows us to decrease the power spectrum errors on $\Omega_m$ and $M_\nu$ by factors of $2.4$ and $3$ (Table \ref{tab:fisher}). The analogous ratios comparing the combination of two marked spectra and the power spectrum of the real-space cold dark matter field were equal to $3.3$ and $4$, respectively, and in this case the improvement is obtained with marked spectra alone. Why is the improvement larger when using cold dark matter? As \cite{Massara_2021} pointed out, marked power spectra are powerful statistics to constrain the neutrino masses when they up-weight low-density regions. In these regions, such as cosmic voids, the $M_\nu-\sigma_8$ degeneracy is likely to be broken --- see \cite{Massara_2015} and their Figure 9 and 13 --- since neutrinos represent a large fraction of the total matter in low-density regions, so that their effects should be more pronounced than in high-density regions, where they are an infinitesimal component of the total matter. The galaxies considered in this work live in high mass halos ($M_h>10^{13}~h^{-1}M_{\odot}$), and the galaxy marked power spectra might not have access to low-density regions in the matter field. This could be the reason why the galaxy marked spectra are not outperforming the galaxy power spectrum as much as the cold dark matter marked spectra.

\subsection{Matter density environment}

\begin{figure}[ht!]
\includegraphics[width=1\columnwidth]{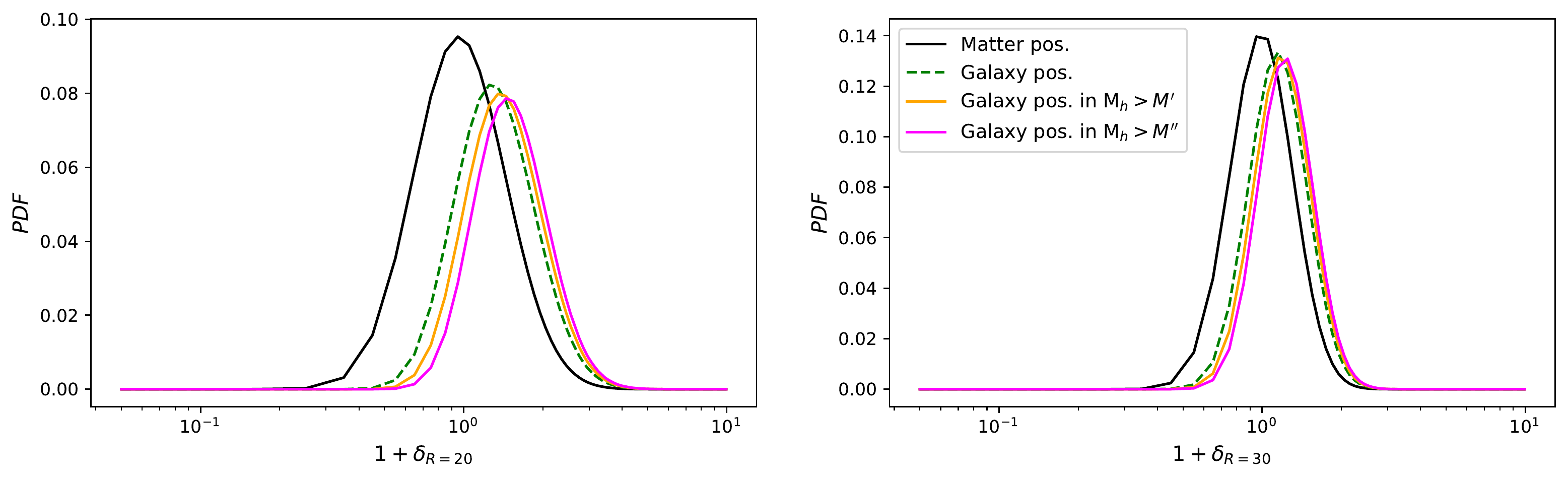}
\caption{\label{fig:density}Probability distribution function of sphere with radius $R=20~h^{-1}$Mpc (left) and $R=30~h^{-1}$Mpc (right) centered at the location of matter particle (black), galaxy (green), and galaxy in halos with mass ${\rm M}_h>{\rm M}'=6.5\cdot 10^{13}h^{-1}{\rm M}_\odot$ (orange) and ${\rm M}_h>{\rm M}''= 10^{14}h^{-1}{\rm M}_\odot$ (magenta) as a function of their overdensity $\delta_R$.}
\end{figure}

In this section we investigate the connection between the galaxies in our simulations and the mass environment where they live. Figure \ref{fig:density} shows the probability distribution function (PDF) of finding an object in a given local density, defined as the cold dark matter density inside a sphere of radius $R$, with $R=20~h^{-1}$Mpc (left panel) and $R=30~h^{-1}$Mpc (right panel); these values correspond to the smoothing scales of $M_1$, $M_3$ and $M_4$, and the measurements have been performed in the fiducial cosmology. The black lines show the PDF at matter particle locations, and the blue lines indicate the PDF at all galaxy positions. The PDF of the matter particles peaks around $1+\delta=1$, while the PDF of the galaxies presents a peak shifted towards higher values. This is expected since galaxies live in halos, which are biased tracers of the cold dark matter field, and the halos in our simulations have positive bias ($b>1$). The left tail of the PDFs indicates the fraction of objects that are in low-density regions. We can see that the fraction of galaxies in matter densities $\delta\sim 0$ are about $60-80\%$ of the number of matter particles in the same local density. Going to lower densities, the number of galaxies decrease very rapidly and it approaches zero when $\delta\sim -0.6$, while the number of matter particles is non zero up to lower densities. This shows that there are no galaxies in very low-density regions and their marked power spectra cannot access the information in them, whilst the cold dark matter particles can. 

The orange and magenta lines in Figure \ref{fig:density} show the PDF of local densities around galaxies that live in a halos with mass larger than $6.5\times 10^{13}$ and $ 10^{14}h^{-1}{\rm M}_{\odot}$, respectively. Their PDFs are shifted towards higher values of $\delta$, with the shift increasing with the halo mass cut, showing that the larger is the minimum halo mass, the larger is the lower accessible density. Thus, marked power spectra are expected to be particularly powerful statistics when applied to dense samples of galaxies that live in low-mass halos. DESI, Euclid, the Roman Space Telescope, and PFS are expected to observe a large number density of galaxies, some of which are hosted by halos with masses smaller than the minimum halo mass in the Quijote simulations that live in low matter density regions.
Thus, we believe the analysis performed in this work should provide conservative bounds, and future spectroscopic surveys will be able to exploit the potential of marked power spectra even further, thanks to observations of galaxies in lower matter density regions.

\subsection{Comparison with galaxy bispectrum}
\cite{Molino_Bk} have performed a Fisher analysis on the Molino catalogs using the monopole of the bispectrum. Their Table 2 reports the marginalized constraints on the cosmological and HOD parameters obtained using the power spectrum (monopole and quadrupole) and the monopole of the bispectrum up to scales $k_{\rm max}=0.5~h~{\rm Mpc}^{-1}$, allowing us to perform a clear comparison with our results in Table~\ref{tab:best_Mk}. The constraints from the redshift-space power spectrum are the same, as expected. The constraints from the monopole of the bispectrum are comparable to the ones obtained by combining the four marked power spectra (monopole and quadrupole) for all cosmological parameters, except $\sigma_8$ and $M_\nu$. The value reported in \cite{Molino_Bk} for the error on $\sigma_8$ is 0.034, which is twice the one obtained with four marked power spectra. This difference is probably due to not including the quadrupole in the bispectrum analysis, which we find crucial to break the galaxy bias-$\sigma_8$ degeneracy in both the power spectrum and in the marked power spectrum analysis. On the other hand, the bispectrum can set tighter constraints on the sum of neutrino masses $M_\nu$ having a Fisher constraint equal to 0.073 eV, which is 1.6 time smaller than our result from the four marked power spectra. The marked power spectrum can be written in terms of higher order statistics using a perturbation theory expansion on large scales \citep[see][]{Philcox_2020,Philcox_2021}. This expansion contain only a subset of the bispectrum configurations, but encompass some configurations of the trispectrum and higher order statistics as well. We expect the marked power spectrum to contain at least part of the information contained in the bispectrum, but not all mark models contain that information. Only mark models that up-weight low density regions can efficiently retrieve non-Gaussian information from the galaxy field (see \cite{Massara_2021} for discussion on marks with $p<0$). As mentioned above, the galaxies in Molino do not sample the very low density regions in the matter field; we suspect this to be the motivation for obtaining a larger error on $M_\nu$ with marked power spectra than with the bispectrum, since the information on neutrinos is mostly in voids and high order point functions (but they are not entirely contained in marked power spectra). Given their different information content, bispectrum and marked power spectra can complement each other and a combined analysis is expect to retrieve more cosmological information.

\section{Conclusions}
\label{sec:conclusions}

In this work we considered galaxy marked power spectra, which are two-point statistics of the galaxy field transformed by weighting each galaxy with a mark. The mark is a function of the local density field, defined at some scale $R$. Different values for the scale $R$ and different functional forms for the mark allow us to generate different marked fields from the same initial one. 

We considered 60 different transformations and used the corresponding marked power spectra to perform a Fisher analysis on the Molino mock galaxy catalog, built via an HOD framework upon the halo catalogs of the Quijote simulations. We selected the four mark models that achieve tighter cosmological constraints but do not exhibit noisy derivatives. The constraints obtained from the monopole and quadrupole of these statistics are compared to the Fisher constraints from the galaxy power spectrum. Each of the marked power spectra can decrease the standard power spectrum constraints on $\Omega_m$, $\Omega_b$, $h$, $n_s$ and $M_\nu$ by $15-25\%$, and on $\sigma_8$ by a factor of 2 when the quadrupole is included in the analysis. The combination of the four marked power spectra and the power spectrum can tighten these constraints even more, bringing them to be $2.5-3.5$ times smaller than the ones obtained with the standard power spectrum alone. The decrease is equal to a factor of $6$ for $\sigma_8$, allowing to reach a $2\%$ accuracy.

This analysis has been performed using a modest volume equal to $1~h^{-3}{\rm Gpc}^{3}$ and a galaxy sample with number density equal to $1.63 \times 10^{-4} h^{3} {\rm Mpc}^{-3}$. On one hand, our results might be optimistic for the considered volume because of noise in the derivatives and the inability to check their convergence by changing the step size in the derivatives with respect to the cosmological parameters. On the other hand, future spectroscopic surveys, such as DESI, Euclid and the Roman Space telescope, will probe much larger cosmic volumes and will observe much denser galaxy populations. This should enable to obtain even tighter constraints on all the cosmological parameters. In particular, being able to trace the inner part of voids with a larger number density of galaxies should enable the marked power spectra that up-weight low density regions to retrieve even more cosmological information than the one reported in this analysis. Moreover, the constraints of this analysis have been obtained without any information coming from the cosmic microwave background (CMB). The inclusion of CMB prior is expected to improve the results of our analysis. 

As a next step, we plan to perform the cosmological analysis with marked power spectra in available galaxy surveys. We are building a simulation-based framework to perform this analysis without 
requiring an analytical model for the marked power spectra and their covariance. This setup could be extended to up-coming surveys and their specifics.

\section{Appendix}
\label{sec:appendix}

We test the stability of our analysis against changes in the number of simulations used to compute derivatives and covariance matrices, and when using different estimators for the derivatives with respect to the neutrino masses. 

\begin{figure}[h!]
\centering
\includegraphics[width=0.9\textwidth]{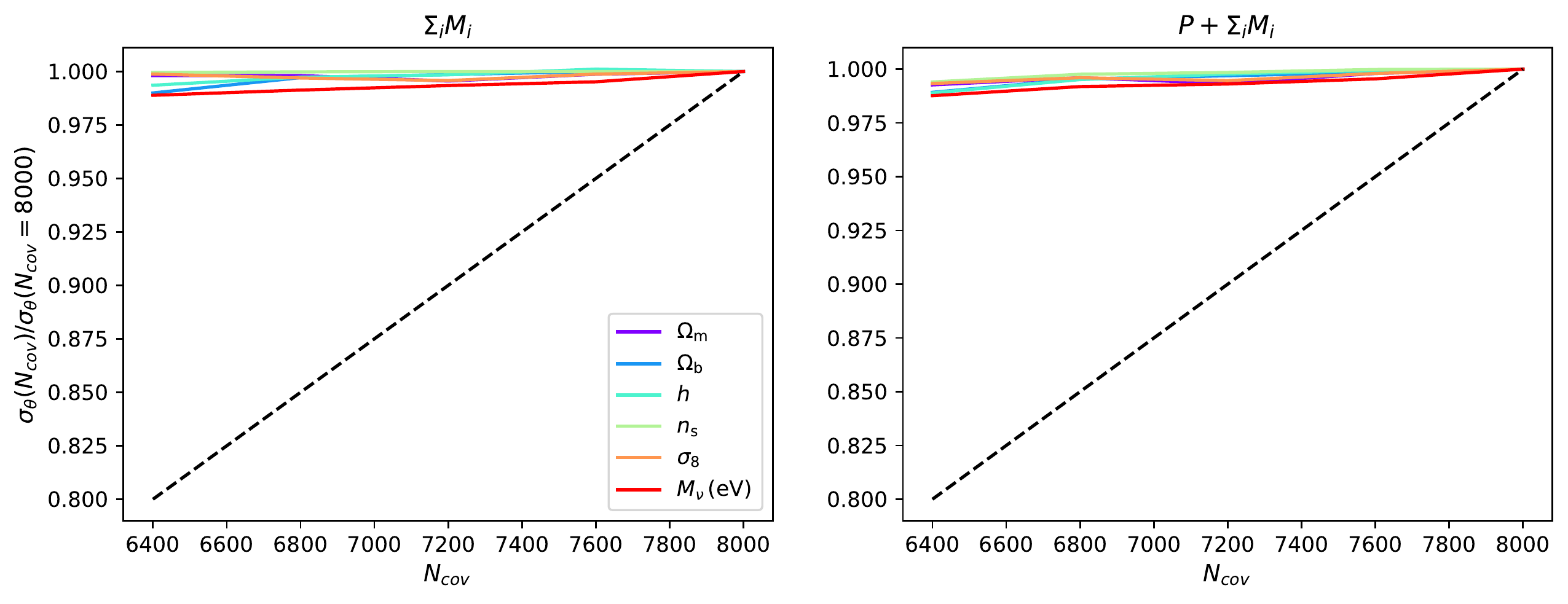}
\caption{Convergence test on the covariance of combination of four marked power spectra (left panel) and of the power spectrum and the four marked power spectra (right panel). The plots show the ratio between the Fisher constraints using the covariance matrix from $N_{\rm cov}$ number of mocks and using the covariance matrix from $8,000$ mocks, as a function of $N_{\rm cov}$. The black dashed lines indicate the variation in number of mocks compare to the full sample, $N_{\rm cov}/8,000$. \label{fig:convCov}}
\end{figure}

Figure~\ref{fig:convCov} displays the marginalized errors on the cosmological parameters as a function of the number of mocks ($N_{cov}$) used to compute the covariance matrix. The errors are normalized by the ones obtained from a covariance matrix computed with 8,000 mocks. The left panel shows the results using as observable the four marked power spectra in Table~\ref{tab:best_Mk}, and the right panel reports the results using all the statistics, i.e. the power spectrum and the four marked power spectra. The black dashed line indicates the change in the number of mocks $N_{cov}$ with respect to 8,000. All the errors are within $2\%$ of the ones obtained with the benchmark $N_{cov}=8,000$.

\begin{figure}[ht!]
\centering
\includegraphics[width=0.9\textwidth]{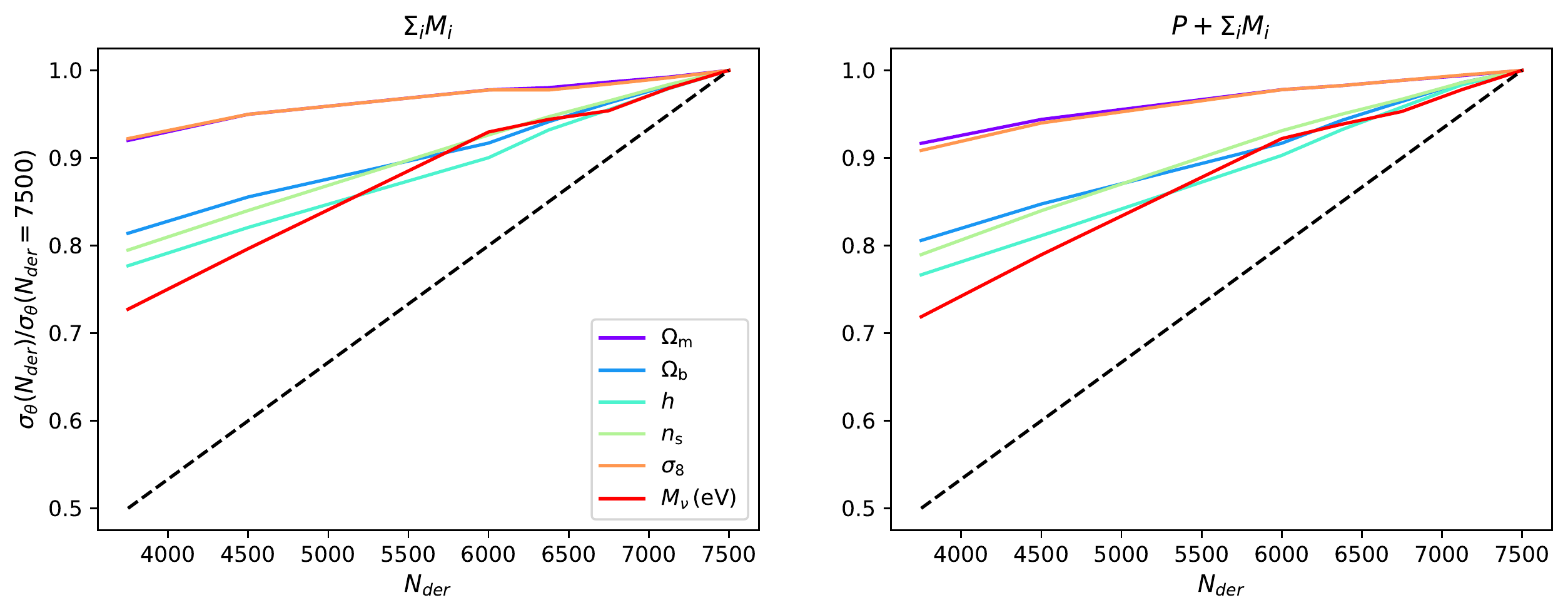}
\caption{Convergence test on the derivatives for combination of four marked power spectra (left panel) and of the power spectrum and the four marked power spectra (right panel). The plots show the ratio between the Fisher constraints using the derivatives from $N_{\rm der}$ number of mocks and from $7,500$ mocks, as a function of $N_{\rm der}$. The black dashed lines indicate the variation in number of mocks compare to the full sample, $N_{\rm der}/7,500$. \label{fig:convDer}}
\end{figure}

Similarly, Figure~\ref{fig:deriv} shows the change in marginalised error as a function of the number of mocks used to compute the derivatives $N_{der}$, and normalized by the benchmark $N_{der}=7,500$. Left and right panels report the results using the four marked spectra and the power spectrum plus the four mark models. For all values of $N_{der}$ the relative variation in the number of mocks (black dashed line) is larger than the relative variation in the marginalized errors (colored lines).

\begin{figure}[h!]
\centering
\includegraphics[width=0.55\textwidth]{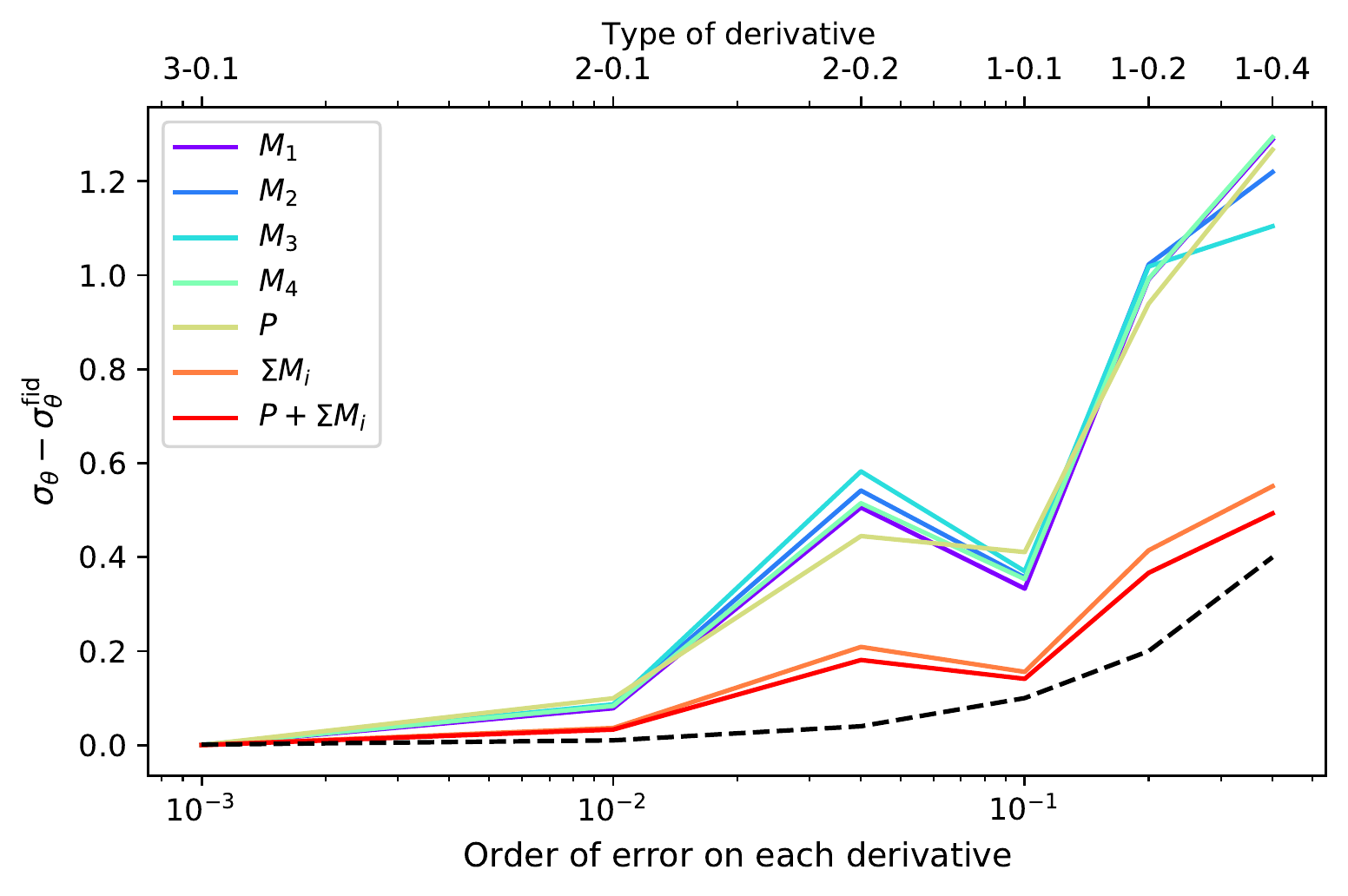}
\caption{Convergence test on different types of $M_\nu$ derivatives for different statistics and different combinations of them. The plots show the difference between the Fisher constraints on $M_\nu$ using one of the derivative estimator in Equation~\ref{eq:deriv_Mnu} and using the fiducial estimator (number 3) employed in our analysis, as a function of the error expected for that estimator (or type of estimator as shown in the upper axis, where the first digit describe the number associated to the estimator and the second float indicates the $dM_\nu$ interval used in the estimator). The black dashed line display the expected order of magnitude for the error made by each estimator. \label{fig:convMnuDer}}
\end{figure}

We also tested how the marginalized errors on the sum of the neutrino masses $M_\nu$ change varying the definition for the derivatives. We consider the estimators in Equation~\ref{eq:deriv_Mnu}, which have an error that scales as powers of $dM_\nu$, which is the step size used to compute the derivatives numerically. Figure~\ref{fig:convMnuDer} displays the magnitude of the error on each estimator as black dashed line, but the actual systematic error could be few times larger. The colored lines show the difference between the Fisher forecast $1-\sigma$ error on $M_\nu$ using a particular estimator and the same quantity obtained with the benchmark estimator (number 4) assumed throughout the paper because it should be the most accurate one. Different colors indicates different statistics considered in the Fisher analysis and all results have been obtained including scales up to $k_{\rm max} = 0.5~h~{\rm Mpc^{-1}}$. The relative error (colored lines) is only few times larger than the systematic error of the estimator (black dashed line) for all statistics when they are considered singularly in the Fisher analysis. When combinations of statistics are considered (orange and red lines), the relative error is very close to the systematic one for each estimator.  

\begin{figure}[h!]
\includegraphics[width=1\columnwidth]{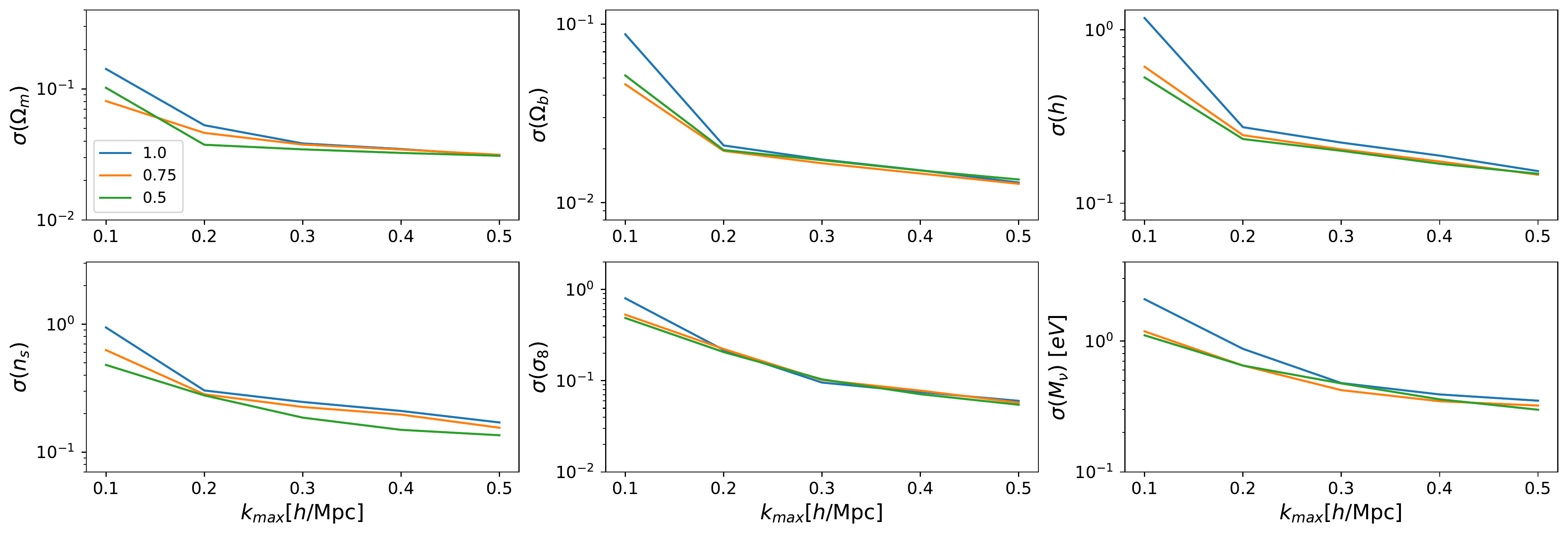}
\caption{Marginalized errors on cosmological parameters for the statistic $M_1$, as a function of the maximum wavenumber considered in the analysis. The different colors correspond to different fraction of subsampled galaxy fields: blue lines display constraints using the full sample, orange and green lines show results using $75\%$ and $50\%$ of the total galaxies, respectively.  \label{fig:errors_subsGal_M1}}
\end{figure}

\begin{figure}[h!]
\includegraphics[width=1\columnwidth]{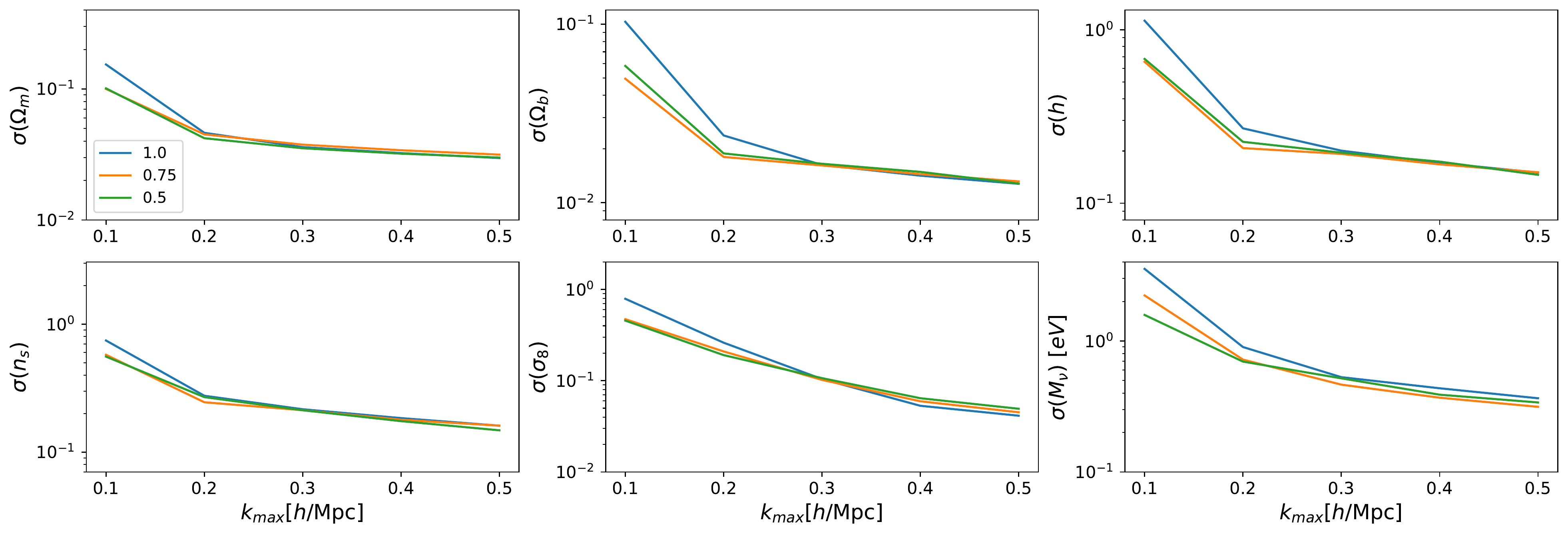}
\caption{Marginalized errors on cosmological parameters for the statistic $M_3$, as a function of the maximum wavenumber considered in the analysis. The different colors correspond to different fractions of subsampled galaxy fields: blue lines display constraints using the full sample, orange and green lines show results using $75\%$ and $50\%$ of the total galaxies, respectively.  \label{fig:errors_subsGal_M3}}
\end{figure}

Lastly, we want to test how the shot-noise affect the estimated constraining power of marked power spectra. Indeed we do not have a good theoretical model for it and cannot subtract it as we did with the power spectrum. We subsampled the galaxy field by randomly selecting $50\%$ and $75\%$ of the galaxies, utilize these subsamples to compute marked power spectra with mark models as in $M_1$ ($R=30h^{-1}$Mpc, $p=1$, $\delta_s=0.1$) and $M_3$ ($R=20h^{-1}$Mpc, $p=1$, $\delta_s=0.5$), and use these measurements to perform the Fisher analysis. Figures~\ref{fig:errors_subsGal_M1} and~\ref{fig:errors_subsGal_M3} show the comparison between the results from the two levels of subsampling and the ones from the full sample, as a function of the maximum wavenumber used in the analysis. We can notice that the marginalized errors on the cosmological parameters are not changed by the number of tracers used to compute the marked power spectrum, specially when small scales are included. Since the only different among them is the number of tracers, and thus the level of shot-noise, we can conclude that our results do not depend on the shot-noise of galaxies and the shot-noise do not need to be subtracted. 

\section*{Acknowledgments}
P.L. acknowledges STFC Consolidated Grant ST/T000473/1. A.M.D. is supported by the Tomalla Foundation for Gravity and the SNSF project ``The  Non-Gaussian  Universe and  Cosmological Symmetries", project number:200020-178787.

\bibliography{mybiblio}{}
\bibliographystyle{aasjournal}



\end{document}